\documentclass[american,aps,pra,reprint,superscriptaddress, longbibliography,floatfix]{revtex4-1}

\usepackage[T1]{fontenc}
\usepackage[latin9]{inputenc}
\usepackage{color}
\usepackage{babel}
\usepackage{amsthm}
\usepackage{amsmath}
\usepackage{amssymb}
\usepackage{graphicx}
\usepackage{calrsfs}
\usepackage[final]{microtype}
\DeclareMathAlphabet{\pazocal}{OMS}{zplm}{m}{n}

\makeatletter
\usepackage{babel}
\usepackage[unicode=true,pdfusetitle, bookmarks=true,bookmarksnumbered=false,bookmarksopen=false,  breaklinks=false,pdfborder={0 0 0},backref=false,colorlinks=false] {hyperref}
\hypersetup{ colorlinks,linkcolor=myurlcolor,citecolor=myurlcolor,urlcolor=myurlcolor}
\usepackage{colortbl}\definecolor{myurlcolor}{rgb}{0,0,0.7}

	\definecolor{armygreen}{rgb}{0.29, 0.33, 0.13}

\usepackage{physics}
\usepackage[normalem]{ulem}

\newcommand*{\Scale}[2][4]{\scalebox{#1}{$#2$}}%

\begin{document}
\title{Entanglement Asymmetry in non-Abelian Anyonic Systems}

\author{Nicetu Tibau Vidal}
\thanks{These authors contributed equally to this work.}
\affiliation{Clarendon Laboratory, University of Oxford, Parks Road, Oxford OX1 3PU, United Kingdom}
\affiliation{QICI Quantum Information and Computation Initiative, Department of Computer Science, The University of Hong Kong, Pok Fu Lam Road, Hong Kong}%
\affiliation{HKU-Oxford Joint Laboratory for Quantum Information and Computation}

\author{Ved Kunte}
\thanks{These authors contributed equally to this work.}
\affiliation{Universit\'{e} Grenoble Alpes, CNRS, Grenoble INP, Institut N\'{e}el, 38000 Grenoble, France}

\author{Lucia Vilchez-Estevez}
\affiliation{Clarendon Laboratory, University of Oxford, Parks Road, Oxford OX1 3PU, United Kingdom}

\author{Mohit Lal Bera}
\affiliation{Departamento de F\'{i}sica Te\'{o}rica and IFIC,  Universidad de Valencia-CSIC, 46100 Burjassot (Valencia), Spain}
\affiliation{ICFO - Institut de Ci\`encies Fot\`oniques, The Barcelona Institute of Science and Technology, 08860 Castelldefels (Barcelona), Spain}

\author{Manabendra Nath Bera}
\email{mnbera@gmail.com}
\affiliation{Department of Physical Sciences, Indian Institute of Science Education and Research (IISER), Mohali, Punjab 140306, India}

\begin{abstract}
Non-Abelian anyons, a promising platform for fault-tolerant topological quantum computation, adhere to the charge super-selection rule (cSSR), which imposes restrictions on physically allowed states and operations. However, the ramifications of cSSR and fusion rules in anyonic quantum information theory remain largely unexplored. In this study, we unveil that the information-theoretic characteristics of anyons diverge fundamentally from those of non-anyonic systems such as qudits, bosons, and fermions and display intricate structures. In bipartite anyonic systems, pure states may have different marginal spectra, and mixed states may contain pure marginal states. More striking is that in a pure entangled state, parties may lack equal access to entanglement. This entanglement asymmetry is manifested in quantum teleportation employing an entangled anyonic state shared between Alice and Bob, where Alice can perfectly teleport unknown quantum information to Bob, but Bob lacks this capability. These traits challenge conventional understanding, necessitating new approaches to characterize quantum information and correlations in anyons. We expect that these distinctive features will also be present in non-Abelian lattice gauge field theories. Our findings significantly advance the understanding of the information-theoretic aspects of anyons and may lead to realizations of quantum communication and cryptographic protocols where one party holds sway over the other.
\end{abstract}
\maketitle

\section{Introduction}
Anyons are 2D topological quasi-particles that follow exotic statistics \cite{Pachos2012IntroductionComputation, Freedman2003TopologicalComputation, Nayak2008Non-AbelianComputation, Simon2023TopologicalQuantum}. Due to their topological nature, anyons are robust against local perturbations, making them a prime candidate for fault-tolerant quantum computation \cite{Kitaev_2001, Kitaev2003Protection, KITAEV2006, Kitaev_2006, Sarma2005}. Unlike distinguishable particles, indistinguishable ones are constrained by super-selection rules. For instance, fermions respect the parity super-selection rule \cite{Wightman1992, Wick1952TheParticles} which plays important roles in the quantum information and computation theory of fermions \cite{Kraus2009PairingPerspective, Friis2013Fermionic-modeInformation, Friis16, Johansson16, TibauVidal2021QuantumFermions, TibauVidal2022AFermions}. The anyonic particles respect the charge super-selection rule (cSSR) \cite{Aharonov1967, Wick1970, Fredenhagen1989SuperselectionTheory}. It states that the superposition of two states with different total topological charges is unphysical and should not be considered in the theory. The cSSR is expected to find deep implications in the quantum information theory of anyons. While studies have been conducted to characterize quantum correlations such as anyonic entanglement and their potential applications \cite{Kitaev_2006, Levin_2006, Bonderson2017AnyonicEntropy, Kato2013}, a comprehensive understanding of how cSSR imposes constraints on anyonic quantum states and operations and consequently affects tasks in information theory, is still lacking. 

This article unveils, for the first time, the distinct information-theoretic characteristics of non-Abelian anyons compared to non-anyonic systems such as qudits, bosons, and fermions. Our investigation delves into the properties of quantum states and operations that respect cSSR, particularly correlations within bipartite anyonic systems. While conventional wisdom assumes equal entanglement sharing between parties in any bipartite system, this is not generally true to anyonic systems. There are pure states $\ket{\psi_{AB}}$ of bipartite anyonic systems $AB$ that exhibit different spectra for the states of $A$ and $B$. There are also pure states that have different entanglement sharing for $A$ and $B$. This is demonstrated by the fact that the fidelity of quantum teleportation from $A$ to $B$ can differ from the fidelity for teleportation from $B$ to $A$. In fact, there are some bipartite anyonic pure states for which only one-way quantum teleportation is possible.

\section{Fibonacci anyons and notations}
In what follows, we focus on Fibonacci anyons \cite{Read1999, Slingerland2001229, Bonesteel2005, Hormozi_2007, Ardonne_2007, Trebst2008}, which provide the simplest model to study non-Abelian anyons and are essential for topological quantum computing. However, our analysis is extendable to any non-Abelian anyon system.  

Fibonacci anyons are described using two types of anyonic (topological) charges, i.e., $a\in\{e,\tau\}$ where $e$ represents the vacuum (or absence of anyon) and $\tau$ is a Fibonacci excitation. Their fusion rules are characterized by $\tau \times e=\tau$, $e \times \tau=\tau$, and $\tau\times \tau=e+\tau$. The global charge after the fusion of two $\tau$ particles must be specified since it gives rise to two orthogonal states \cite{Pachos2012IntroductionComputation, Simon2023TopologicalQuantum, Trebst2008, Nayak2008Non-AbelianComputation}. A canonical basis for multiple Fibonacci anyons is specified by their fusion order, and different canonical bases are interrelated via $F$-moves (see Appendix~\ref{sec:appendixa}) \cite{Pachos2012IntroductionComputation, Simon2023TopologicalQuantum, Bonderson2007Non-AbelianPh.D., NicetuThesis}. In subsequent discussions, the term `anyons' exclusively denotes Fibonacci anyons.

The anyonic states and bases are conventionally expressed in terms of diagrams. For convenience, we denote them using Dirac notation. The correspondence between the diagrams and Dirac notations is presented in Fig.~\ref{fig:DiracDiagram}(x). See Appendix~\ref{sec:appendixa} for more details. For a bipartite system with two anyons, the state space is spanned by five orthonormal states; $\{ \ket{e,e;e}, \ket{\tau,\tau; e}\}$ for the sector with global charge $e$ and $\{\ket{\tau,e;\tau}, \ket{e,\tau;\tau}, \ket{\tau,\tau; \tau}\}$ for the sector of global charge $\tau$ \cite{Pachos2012IntroductionComputation}. For a system with four anyons with the partitions $(A_1,A_2)$ and $(B_1,B_2)$, there are $34$ orthonormal states denoted by $\{ \ket{(a_1,a_2)(b_1,b_2);a, b;g} \}$, where $a_i$ and $b_i$ are the states of the anyons $A_i$ and $B_i$ respectively; $a,b$ are the charges after fusions of anyons $a_1,a_2$ and anyons $b_1,b_2$ respectively; and $g$ is the global charge after fusion of anyons $a,b$. The anyons respect cSSR, which states that every physical state can only have superpositions between states with the same global charge \cite{Aharonov1967, Wick1970, Fredenhagen1989SuperselectionTheory}. Thus, all observables and unitaries must be block-diagonal in the (global) charge sectors.

\begin{figure}
\centering    
\includegraphics[width=8.5cm,height=4cm]{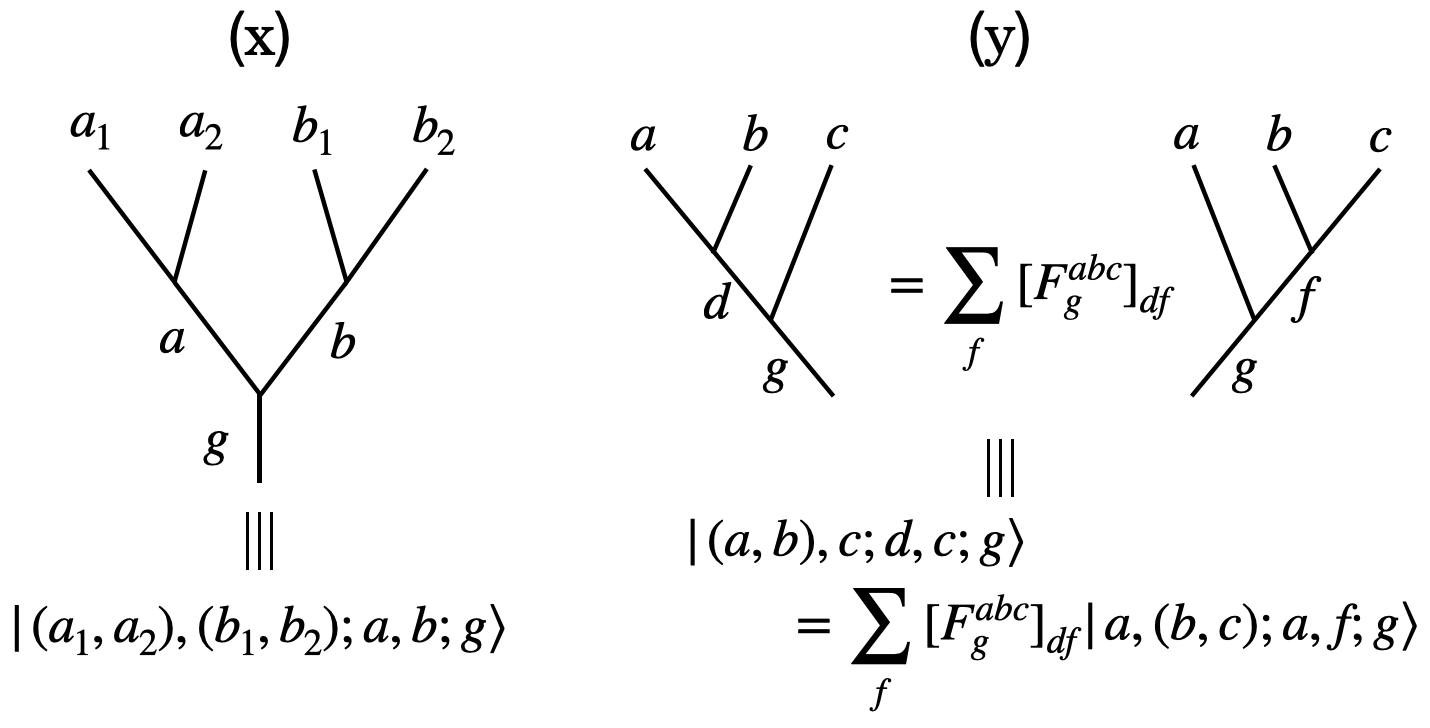}
\caption{Correspondence between Dirac and diagrammatic notations of anyons. (x) The figure on the left represents how diagrammatic notations for anyons are related to Dirac notation. (y) The figure on the right shows the transformation between (non-associative) anyonic basis via $F$-moves, both in diagrammatic and Dirac notations. In both figures, time is flowing upward.
\label{fig:DiracDiagram}}
\end{figure}

Consider a bipartite system $AB$, where $A$ is a party with anyons $A_1, A_2$ and $B$ is a party with $B_1, B_2$. Then, a cSSR-respecting local operator $\ketbra{a_1,a_2; a}{a'_1,a'_2; a}$ acting on $A$ can be extended to the corresponding global operator acting on the space of $AB$
\begin{align*}
\sum\limits_{b_1,b_2,b,g} \ketbra{(a_1,a_2)(b_1,b_2);a,b;g}{(a'_1,a'_2)(b_1,b_2);a,b;g},
\end{align*}
where the sum is over the values of the labels allowed by the fusion rules. Note that the extended operator also respects cSSR. The above extension can be interpreted as the analogue of the embedding $\ketbra{i}{j} \to \sum_k \ketbra{i k}{j k}$ for qudit systems, where $\{\ket{k}\}$ represents an orthonormal basis spanning the ancillary space. For instance, consider a unitary $U_{A_1}=\text{diag}\{e^{i\phi}, e^{i \eta} \}$ acting on a $1$-anyon system $A_1$ with the (ordered) basis $\{\ket{e}, \ket{\tau} \}$. Then, the extension of this unitary $U_{A_1A_2}$ acting on $2$-anyon system $A_1A_2$ is given by $U_{A_1A_2}=\text{diag}\{e^{i \phi}, e^{i \eta}, e^{i \eta}, e^{i \phi}, e^{i \eta} \}$ for the (ordered) basis $\{\ket{e,e;e}, \ket{\tau,\tau;e}, \ket{\tau,e;\tau}, \ket{e,\tau;\tau},  \ket{\tau,\tau;\tau} \}$. In the same vein, partial tracing the degree of freedoms of $B$ for an operator $\ketbra{(a_1,a_2)(b_1,b_2); a,b;g}{(a'_1,a'_2)(b'_1,b'_2);a',b';g}$ acting on $AB$ is done by 
\begin{align}
\Tr_B(\ketbra{(a_1,a_2)(b_1,b_2);a,b;g}{(a'_1,a'_2)(b'_1,b'_2);a',b';g}) \nonumber \\
= \delta_{b_1 b'_1} \delta_{b_2 b'_2} \delta_{b b'} \delta_{a a'} \ketbra{a_1, a_2; a}{a'_1, a'_2; a}, \label{eq:pt2}   
\end{align}
where $\delta_{a a'}$ appears as the consequence of cSSR. For a $2$-anyon system $A_1A_2$, the partial tracing of $A_2$ for an operator $\ketbra{a_1,a_2;g}{a_1',a_2';g}$ becomes $\Tr_B\left(\ketbra{a_1,a_2;g}{a_1',a_2';g}\right) = \delta_{a_2 a_2'} \delta_{a_1 a_1'} \ketbra{a_1}$. The extension and partial tracing are generalized to arbitrary operators acting on an $N$-anyon system \cite{Simon2023TopologicalQuantum, Bonderson2007Non-AbelianPh.D.}. The notion of correlations that such subsystem identification provides is analogous to the \emph{mode} picture in the fermionic and bosonic literature \cite{Gigena2017BipartiteSystems,Gigena2015EntanglementSystems,Banuls09,bosonicinfo,DAriano2014FermionicEntanglement,Perinotti2021ShannonFermions,TibauVidal2023CreationAnyons}.

\section{Anyonic states and marginal spectra ambiguity}
We start with correlations in anyonic pure states. In previous works \cite{Bonderson2017AnyonicEntropy, Bonderson2007Non-AbelianPh.D.}, it is considered that any anyonic state with a (topological) charge $\tau$ cannot be a pure state, e.g., a state $\ketbra{\tau}$ would not be a pure state. We argue that such characterization arises only from considering the isotopy normalization instead of the physical normalization \cite{Simon2023TopologicalQuantum}. Instead, we rely on the information-theory-based definition of pure states \cite{gptframework} that cannot be expressed as a non-trivial probabilistic mixture of two other states. Therefore, any physical state on the boundary of the convex hull is a pure state. Such a definition implies that the pure states are all the states of the form $\rho=\ketbra{\psi}$. Thus, $\ketbra{\tau}$ is a pure state. See Appendix \ref{sec:appendix2} for more details.

Consider a cSSR respecting $2$-anyon bipartite pure state $\rho_{AB}=\ketbra{\psi}_{AB}$, with
\begin{align}
\ket{\psi}_{AB} &=
\frac{1}{\sqrt{2}}\left(\ket{e,\tau;\tau} + \ket{\tau, \tau;\tau}\right).\label{eq:stateasym}
\end{align}
The local density operators of subsystems $A$ and $B$, following anyonic partial trace prescribed in Eq.~\eqref{eq:pt2}, $\rho_A=\frac{1}{2}\left(\ketbra{e}+\ketbra{\tau}\right)$ and $\rho_B=\ketbra{\tau}$ respectively. The disparity between the spectra of the two marginals is evident. While the reduced state of $B$ is pure, the state is maximally mixed for $A$. This observation is distinctly different from non-anyonic systems and is attributed to cSSR and anyonic fusion rules. Due to the non-anyonic Schmidt decompositions of bipartite pure states, the marginal density operators have the same spectra. This is why functions of the marginal spectra are often used to characterize bipartite entanglement systems composed of non-anyonic systems \cite{Nielsen00, bosonicinfo, Friis16, TibauVidal2021QuantumFermions}. This also justifies why the entanglement sharing between two parties is the same. On the contrary, due to the ambiguity, it is clear that a function of marginal spectra cannot characterize the anyonic entanglement. This raises questions on the correctness of quantifying entanglement in bipartite pure anyonic states using entanglement entropy, as done in \cite{Bonderson2017AnyonicEntropy}.

Beyond this, mixed anyonic states exhibit ambiguity between global and marginal spectra. For instance, consider a mixed state $\rho_{AB}=\frac{1}{2} \left(\ketbra{\tau,\tau;e} +\ketbra{\tau,\tau;\tau}\right)$. After partial tracing, the corresponding marginal states become $\rho_A=\ketbra{\tau}$ and $\rho_B=\ketbra{\tau}$. Here, while the global state is mixed, the marginal states are pure. Therefore, having pure marginals does not guarantee that the global state is pure. This is distinctly different from what we see in non-anyonic systems and, yet again, is attributed to anyonic fusion rules and cSSR.

\textit{Correlated anyonic states}---The above observations indicate that the conventional approaches to characterize correlations cannot be applied to anyonic systems. We resort to the primitive definition of uncorrelated state, which states that a state $\rho_{AB}$ of a bipartite system $AB$ is uncorrelated if  
\begin{align}
    \Tr\left(\hat{O}_A \ \hat{O}_B \ \rho_{AB}\right)=\Tr\left(\hat{O}_A \  \rho_A\right)\Tr\left(\hat{O}_B \ \rho_B\right) \label{eq:uncorr}
\end{align}
for any local observables $\hat{O}_A$ and $\hat{O}_B$ belonging to $A$ and $B$ respectively. Using this definition, we proceed to characterize the correlated anyonic states.

For a $2$-anyon system $AB$, the cSSR respecting pure states are:
\begin{align}
   \label{eq:estates} \ket{\psi_e}_{AB}=\alpha_e \ket{e,e;e}+\beta_e\ket{\tau,\tau; e}, \\ \label{eq:taustates} \ket{\psi_\tau}_{AB}=\alpha_\tau \ket{\tau,e;\tau}+\beta_\tau \ket{e,\tau; \tau}+\gamma_\tau \ket{\tau,\tau;\tau},
\end{align}
where $|\alpha_e|^2 + |\beta_e|^2=1$ and $|\alpha_\tau|^2 + |\beta_\tau|^2 + |\gamma_\tau|^2=1$. As per Eq.~\eqref{eq:uncorr}, the state $\ket{\psi_e}_{AB}$ with global charge $e$ is uncorrelated if either $\alpha_e=0$ or $\beta_e=0$. 

There are infinitely many uncorrelated states with global charge $\tau$ assuming the form $\ket{\psi_\tau}_{AB}$. These uncorrelated states are categorized into two classes; one with $\alpha_\tau=0$ and the other with $\beta_\tau=0$. In other words, contrary to their appearance, the states with the structures $\alpha_\tau\ket{e,\tau;\tau}+\gamma_\tau \ket{\tau,\tau;\tau}$ and $\beta_\tau\ket{\tau,e;\tau}+\gamma_\tau \ket{\tau,\tau;\tau}$ are the uncorrelated states. These states satisfy the relation~\eqref{eq:uncorr}, as shown in Appendix~\ref{sec:appendixuncorr}. Note, the state considered in Eq.~\eqref{eq:stateasym}, exhibiting unequal marginal spectra, is an uncorrelated state of the second kind ($\beta_\tau=0$). Having two classes of uncorrelated states intersecting at a point is a unique feature of any non-Abelian anyon theory. Further, we find that the group of products of local unitaries $\pazocal{G}_{AB}=\{\hat{U}_A \hat{V}_B\}$ cannot map any pure uncorrelated state to any other pure uncorrelated state. Instead, $\pazocal{G}_{AB}$ can, at most, change the relative phases of the complex coefficients. 

For pure states, an entangled state is defined to be a state that is not uncorrelated \cite{Nielsen00}. One may find `maximally' entangled states for a fixed global charge for which the marginal states are maximally mixed. However, for a $2$-anyon system, the only maximally entangled states with the global charge $\tau$ are $\frac{1}{\sqrt{2}} \left(\ket{e,\tau;\tau}+e^{i\varphi} \ket{\tau,e;\tau}\right)$. From this, it is clear that, unlike non-anyonic systems, we do not have a basis of maximally entangled states spanning the entire bipartite anyonic space. In addition to these particularities, the bipartite entanglements in anyonic systems are also asymmetric, which we discuss next.     

\section{Quantum teleportation and asymmetric entanglement sharing}
It is commonly known that bipartite entanglement is symmetric with respect to individual parties. For any bipartite entangled state shared between Alice and Bob, any quantum task that can be realized by implementing a one-way local operation and classical communication (LOCC) from Alice to Bob can also be realized using one-way LOCC from Bob to Alice. Remarkably, this is not true in general for anyonic systems, as demonstrated below.

In quantum teleportation, an unknown quantum message, encoded in a quantum state, is teleported using one-way LOCC and an entangled state, where the latter is considered as the resource \cite{Bennett93}. Since a (Fibonacci) $1$-anyon state can only encode a classical bit, we investigate the teleportation of a pure qubit encoded in a pure state $\ket{\varphi_M}$ of a $2$-anyon system ($M$). We may also refer message $\ket{\varphi_M}$ as $(\alpha,\beta)$. The teleportation of $\ket{\varphi_M}$ between Alice and Bob is done exploiting a $4$-anyon state $\ket{R_{AB}}$ shared between them, where
\begin{align*}
    \ket{\varphi_M} &= \alpha\ket{\tau,e;\tau} + \beta\ket{e,\tau;\tau}  \\
    \ket{R_{AB}} &= \Scale[0.97]{\frac{1}{\sqrt{2}}\Big(\ket{(e,e),(e,\tau);e,\tau;\tau} + \ket{(\tau,e),(\tau,e);\tau,\tau;\tau}\Big) }
\end{align*} 
Note, $\ket{R_{AB}}$ is a bipartite entangled state with identical marginal spectra, where the parties $A$ and $B$ have two anyons each. 

Let us now consider the teleportation from Alice to Bob. The global state of $M(AB)$, using the fusion channel resulting in the global charge $e$, is given by
\begin{align*}
    \ket{\psi_{M(AB)}}=&\frac{1}{\sqrt{2}} \Big( \alpha \ket{(\tau,e),(e,e),(e,\tau);\tau,(e,\tau);\tau,\tau;e} \nonumber\\ 
     & +\alpha \ket{(\tau,e),(\tau,e),(\tau,e);\tau,(\tau,\tau);\tau,\tau;e} \nonumber \\
     & +\beta \ket{(e,\tau),(e,e),(e,\tau);\tau,(e,\tau);\tau,\tau;e} \nonumber \\
     & + \beta \ket{(e,\tau),(\tau,e),(\tau,e);\tau,(\tau,\tau);\tau,\tau;e} \Big) 
\end{align*}
As in the usual teleportation protocol, Alice performs a projective-valued measurement (PVM) on the joint system $MA$. She then sends the measurement outcome to Bob via a classical communication channel, and based on that, he performs a local transformation on his system $B$ to retrieve the quantum message. Note that anyonic states are not associative, and recombination of the basis is done using $F$-moves \cite{Pachos2012IntroductionComputation, Simon2023TopologicalQuantum,Nayak2008Non-AbelianComputation,Bonderson2007Non-AbelianPh.D.}, as shown in Fig.~\ref{fig:DiracDiagram}(y). However, the $F$-moves become trivial for states with global charge of $\ket{\psi_{M(AB)}}$ is $e$. Using this, we re-express the state $\ket{\psi_{M(AB)}}$ with $\ket{\psi_{(MA)B}}$, given by 
\begin{align}
    \ket{\psi_{(MA)B}}&=\frac{1}{\sqrt{2}}\Big(\alpha \ket{(\tau,e),(e,e),(e,\tau);(\tau,e),\tau;\tau,\tau;e} \nonumber\\ & +\alpha \ket{(\tau,e),(\tau,e),(\tau,e);(\tau,\tau),\tau;\tau,\tau;e} \nonumber \\
     & +\beta \ket{(e,\tau),(e,e),(e,\tau);(\tau,e),\tau;\tau,\tau;e} \nonumber \\
      & +\beta \ket{(e,\tau),(\tau,e),(\tau,e);(\tau,\tau),\tau;\tau,\tau;e} \Big), \label{eq:asymMAB}
\end{align}
which allows us to easily calculate the outcomes of Alice's measurements on $MA$.  The state again can be recast as
\begin{align}
    \ket{\psi_{(MA)B}}=\frac{1}{2}\Big(&\ket{\lambda_+}\big(\alpha \ket{e,\tau;\tau} +\beta \ket{\tau,e;\tau}\big) \nonumber\\ 
    + &\ket{\lambda_-} \big(\alpha \ket{e,\tau;\tau} -\beta \ket{\tau,e;\tau}\big)   \nonumber\\ 
    +&\ket{\eta_+} \big(\alpha \ket{\tau,e;\tau} +\beta \ket{e,\tau;\tau}\big)   \nonumber\\   +&\ket{\eta_-} \big(\alpha \ket{\tau,e;\tau} -\beta \ket{e,\tau;\tau}\big) \Big) \label{eq:teleportY}
\end{align}
where the orthonormal states
\begin{align*}
     \text{\small{$\ket{\lambda_{\pm}}= \frac{1}{\sqrt{2}} \Big( \ket{(\tau,e),(e,e);\tau,e;\tau} \pm  \ket{(e,\tau),(\tau,e);\tau,\tau;\tau} \Big)$}}, \nonumber \\
     \text{\small{$\ket{\eta_{\pm}}= \frac{1}{\sqrt{2}} \Big( \ket{(\tau,e),(\tau,e);\tau,\tau;\tau} \pm  \ket{(e,\tau),(e,e);\tau,e;\tau} \Big)$}},
\end{align*}
represent $4$-anyon states of $MA$ accessible to Alice. The global charges $\tau$ of $\ket{\lambda_{\pm}}$ and $\ket{\eta_{\pm}}$ fuse with Bob's global charges $\tau$ through the $\tau \times \tau = e$ fusion channel.

It is straightforward to check that Alice can choose the cSSR-allowed measurements applying the projectors $\{\ketbra{\lambda_+}, \ketbra{\lambda_-}, \ketbra{\eta_+},\ketbra{\eta_-} \}$; sends classical information to Bob whenever there is a click in the apparatus due to projections; Bob then applies (cSSR-allowed) encoded Pauli unitaries $\{X, Y, I, Z\}$ on his system respectively. Here, we denote $Z=\ketbra{\tau,e;\tau}-\ketbra{e,\tau;\tau}$, and accordingly $X, Y$, and identity operator $I$. After the operations, the reduced state of Bob becomes $\ket{\varphi_M}$, and with this, a perfect (with unit fidelity) teleportation of the quantum message $(\alpha, \beta)$ from Alice to Bob is realized.

Now, we investigate quantum teleportation in the reverse direction and show that Bob cannot perfectly teleport the quantum message to Alice. In this case, the message state $\ket{\varphi_M}$ is attached with the resource state $\ket{R_{AB}}$ from Bob's side. The global state of $(AB)M$ is then
\begin{align}
    \ket{\xi_{(AB)M}}&=\frac{1}{\sqrt{2}}\Big( \alpha \ket{((e,e),(e,\tau),(\tau,e);(e,\tau),\tau;\tau,\tau;e} \nonumber\\ 
    & +\alpha \ket{(\tau,e),(\tau,e), (\tau,e);(\tau,\tau), \tau;\tau,\tau;e}  \nonumber \\
    & +\beta \ket{(e,e),(e,\tau), (e,\tau);(e,\tau), \tau;\tau,\tau;e}  \nonumber \\
    & +\beta \ket{(\tau,e),(\tau,e), (e,\tau);(\tau,\tau), \tau;\tau,\tau;e} \Big). \label{eq:asymABM}
\end{align}
Again, using the (trivial) $F$-moves due to the global charge $e$, we re-express the state 
\begin{align}
    \ket{\xi_{A(BM)}}&=\frac{1}{\sqrt{2}}\Big( \alpha \ket{((e,e),(e,\tau),(\tau,e);e,(\tau,\tau);e,e;e} \nonumber\\ 
    & +\alpha \ket{(\tau,e),(\tau,e), (\tau,e);\tau,(\tau, \tau);\tau,\tau;e} \nonumber \\
    & +\beta \ket{(e,e),(e,\tau), (e,\tau);e,(\tau, \tau);e,e;e} \nonumber \\
    & + \beta \ket{(\tau,e),(\tau,e), (e,\tau);\tau,(\tau, \tau);\tau,\tau;e} \Big), \label{eq:asymABMM}
\end{align}
where $BM$ is combined together, and Bob can apply local operations on them. One can immediately notice the difference between Eq.~\eqref{eq:asymABMM} and Eq.~\eqref{eq:asymMAB}. In Eq.~\eqref{eq:asymMAB}, all the local states of $MA$ in superposition have identical charge $\tau$. 

In contrast, in Eq.~\eqref{eq:asymABMM}, the local states of $BM$ in the first and third terms in the superposition have a global charge $e$, while the second and the fourth terms have a global charge $\tau$. Due to the anyonic cSSR, superpositions of states with different global charges are not allowed. Therefore, a local measurement using arbitrary projectors is not possible, e.g., projectors acting on a subspace of $BM$ that allows superposition between global charges $e$ and $\tau$ are not physical.

Due to this fundamental constraint attributed to cSSR, Bob can only perform measurements on $BM$ with non-trivial projectors that either act on the subspace spanned by $\{\ket{(e,\tau),(\tau,e);\tau,\tau;e}, \ket{(e,\tau),(e,\tau);\tau,\tau;e}\}$ with global charge $e$ or subspace spanned by $\{\ket{(\tau,e),(\tau,e);\tau,\tau;\tau}, \ket{(\tau,e),(e,\tau);\tau,\tau;\tau} \}$ with global charge $\tau$. Regardless of any choice of these restricted set of measurements on the state $\ket{\xi_{A(BM)}}$, the reduced state of Alice can at most be a probabilistic mixture of $\ketbra{e,e;e}$ and $\ketbra{\tau,e;\tau}$. The quantum message $(\alpha, \beta)$ encoded in $\ket{\varphi_M}$ can never be teleported to Alice after the teleportation protocol is completed. 
Note, if one ignores the cSSR and allows arbitrary measurements on $BM$ by Bob, it is possible to implement a perfect quantum teleportation from Bob to Alice.

From the above, it is clear that while Alice can perfectly teleport to Bob, Bob cannot perfectly teleport an unknown quantum message to Alice. This asymmetry in teleportation is not particular to the choice of the anyonic entangled state $\ket{R_{AB}}$. It is rather a generic feature and can be shown in other bipartite pure entangled states (see Appendix~\ref{sec:tpexamples}). As the ability to quantum teleport is attributed to the presence of quantum entanglement in the resource state, the asymmetry in quantum teleportation signifies that entanglement sharing among the parties is not equal. 

\section{Conclusions}
As indicated above, the cSSR and fusion rules make the information theory for (non-Abelian) anyons fundamentally different from non-anyonic systems. It prohibits coherent superposition between states with different topological charges, thereby imposing restrictions on physically allowed anyonic states, operations, and information processing. The correlations in anyonic systems are distinctively different from non-anyonic ones and display richer structure. A bipartite pure anyonic state can have local states with different marginal spectra, and a mixed bipartite state may have pure local states. What is more striking is that, for a bipartite pure entangled state, the parties do not have uniform access to entanglement. This asymmetry in entanglement is manifested in quantum teleportation, where only one party can perfectly teleport a quantum message to another party rendering the latter having no access to entanglement. These defy common understanding and indicate that the conventional approach in characterizing information and correlation falls short for non-Abelian anyonic systems. We expect that these distinctive traits to be present in non-Abelian lattice gauge field theories \cite{bianchi2024nonabelian, Donnelly:2011, Ghosh:2015, Aoki2015OnTD, Casini14, Casini:2019, Klco2021}, as well, since the local physical Dirac observables are similarly constrained by the fusion rules and cSSR. 

Our analysis and results warrant a newer understanding of information-theoretic properties of anyons and non-Abelian lattice gauge fields and are expected to open up a field of research exploring fundamental properties of anyonic quantum correlations and their manipulation. On the applied level, with asymmetric access to bipartite entanglement, quantum tasks such as communication or cryptographic protocols may be devised where one party has the superiority in accessing correlation and manipulation of information over the other.  

\section{Acknowledgements}
N.T.V. gratefully acknowledges financial support from the John Templeton Foundation through grant 62312, The Quantum Information Structure of Spacetime (qiss.fr). The opinions expressed in this publication are those of the authors and do not necessarily reflect the views of the John Templeton Foundation. V.K. would like to thank the MSCA Cofund QuanG (Grant Number: 101081458) funded by the European Union. M.L.B. acknowledges financial support from the Spanish MCIN/AEI/10.13039/501100011033 grant PID2020-113334GB-I00, Generalitat Valenciana grant CIPROM/2022/66, the Ministry of Economic Affairs and Digital Transformation of the Spanish Government through the QUANTUM ENIA project call - QUANTUM SPAIN
project, and by the European Union through the Recovery, Transformation and Resilience Plan - NextGenerationEU
within the framework of the Digital Spain 2026 Agenda, and by the CSIC Interdisciplinary Thematic Platform (PTI+)
on Quantum Technologies (PTI-QTEP+). This project has also received funding from the European Union's Horizon
2020 research and innovation program under grant agreement CaLIGOLA MSCA-2021-SE-01-101086123. Views and opinions expressed are, however, those of the author(s) only and do not necessarily reflect those of the European Union or QuanG. Neither the European Union nor the granting authority can be held responsible for them. L.V.E is supported by the Clarendon Fund.

\bibliography{AQIT}

\newpage

\appendix 

\onecolumngrid

\section{Basic anyonic structures: Notations, fusion, braiding, and anyon diagrams}
\label{sec:appendixa}

In this section, we introduce the basic notions of the theory of non-Abelian anyons. Anyons are a type of quasi-particle that exists in two-dimensional spaces. They can be seen as a constrained quantum system from the quantum information perspective. We introduce the basic formalism, and we give the equivalence between the bra-ket notation \cite{Pachos2012IntroductionComputation}, we use in the main text, and the diagrammatic notation used in the literature \cite{Simon2023TopologicalQuantum, Bonderson2007Non-AbelianPh.D., Bonderson2017AnyonicEntropy, Trebst2008}.   

\subsection[Fusion, splitting, $F$-moves and braiding]{Fusion, splitting, F-moves and braiding}
\label{subsec:fusion}

There are several anyon theories. To fix an anyon theory, one needs to fix the particle types $a,b,c,\dots$, the fusion rules, and the $F$ and $R$ matrices. We work with Fibonacci anyons, described by the simplest non-Abelian anyon theory. Fibonacci anyons only have two particle types, $e$ and $\tau$.   

The most striking structural property of anyons is how they are composed. Two anyons can be put together to create a new particle. This process is known as \emph{fusion}. Two particle types $a$ and $b$ can be fused together to produce the particle type $c$. We can describe this process by writing $a \times b = b \times a = c$. However, remarkably, it is possible that two anyons can fuse to multiple types of particles; in this case, we can write: 
\begin{equation}
   \Scale[1.2]{ a\times b =b \times a = \sum_c N_{ab}^c c}
    \label{eq:fusion}
\end{equation}
where $N_{ab}^c$ are the fusion multiplicities, they indicate the number of ways in which $a$ and $b$ can fuse to $c$. Without loss of generality, we consider $N_{ab}^c \leq 1$.

For any anyon theory, there is a trivial anyon $e$ called the vacuum or the identity. This particle type satisfies the property $N_{e a}^b=\delta_{ab}$. Every particle type $a$ also has its own antiparticle $\bar{a}$ such that $N_{ab}^e=\delta_{b \bar{a}}$, so that $\bar{a}$ is the only particle type such that when fuses with $a$ can give the vacuum charge $e$. For Fibonacci anyons, the fusion rules can be written as:
\begin{equation}
  \Scale[1.2]{  e\times e = e, \medspace \medspace \medspace e\times \tau=\tau, \medspace \medspace \medspace \tau\times \tau= e+\tau}
\end{equation}

\begin{figure}[ht]
    \centering
    \includegraphics[width=0.32\textwidth]{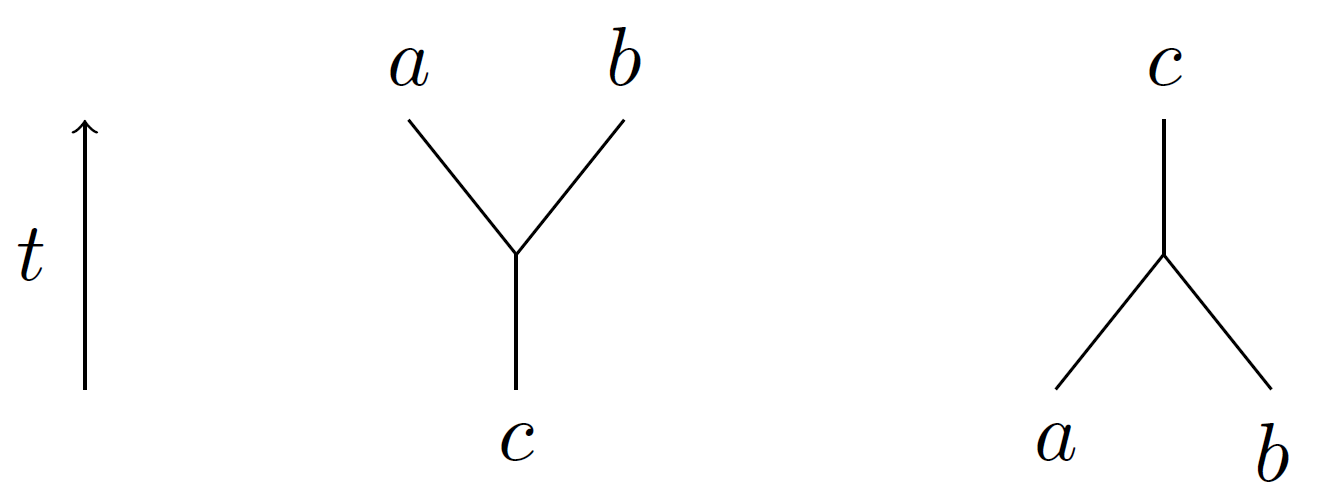}
    \caption{Splitting (left) of one anyon $c$ into $a$ and $b$, and fusion (right) of $a$ and $b$ into $c$.}
    \label{fig:fusion}
\end{figure}

The reverse process (going from one particle to two) is called \emph{splitting}. Both processes are shown diagrammatically in Fig.~\ref{fig:fusion} (see also \cite{Bonderson2007Non-AbelianPh.D.}). The time direction is up, and all anyon particles move forward in time.

For $N$-particle anyonic systems, the kinematical or unrestricted Hilbert space $\pazocal{H}_N$ is given by considering each splitting diagram as an orthonormal vector. Thus, $\pazocal{H}_2$ is spanned by the set of states $\{ \ket{a,b;c}  | N_{ab}^c=1\}$ that denote the different splittings, where $a,b,c \in \{e, \tau \}$. For systems with a single anyon particle, the Hilbert space $\pazocal{H}_1$ is given by the span of the particle types interpreted as orthonormal vectors. $\pazocal{H}_1=\langle \{\ket{a}\} \rangle$.   

\begin{figure}[ht]
    \centering
    \includegraphics[width=0.5\linewidth]{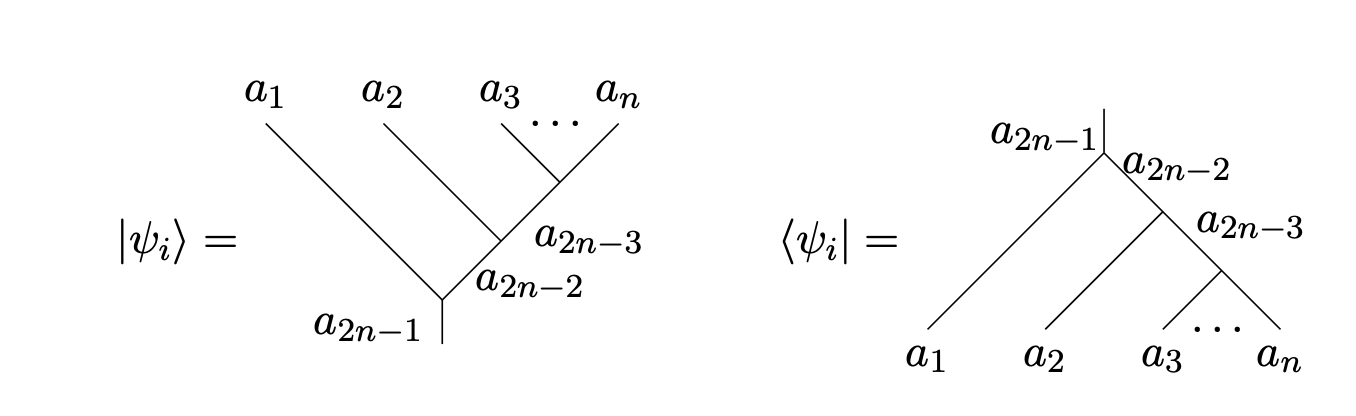}
    \caption{Elements $\ket{\psi_i}$ of orthonormal basis of an anyonic $n$-particle system, given in Eq.~\eqref{eq:ketpachnotation}, and their duals.}
    \label{fig:ket}
\end{figure}

For an anyonic $N$-particle system, the Hilbert space is spanned by the splitting diagrams shown in Fig. \ref{fig:ket}, interpreted as orthonormal vectors. Notice that, since two anyons can fuse to different possibilities, $\dim(\pazocal{H}_{N}) \neq d^N$. For Fibonacci anyons one can easily calculate that $\dim(\pazocal{H}_1)=2$, $\dim(\pazocal{H}_2)=5$, $\dim(\pazocal{H}_3)=13$,\dots  In general, $\dim(\pazocal{H}_N)=F_{2N+1}$, where $F_n$ is the $n$'th Fibonacci number. Therefore, non-Abelian anyon theories and Fibonacci anyons, in particular, have no tensor product structure. 

The states of Fig.~\ref{fig:ket} can be represented with the Dirac notation we use in the main text as 
\begin{equation}\label{eq:ketpachnotation}
\Scale[1.05]{\ket{\psi_i}=\ket{a_1,(a_2, \dots (a_{n-2},(a_{n-1}, a_n))\dots); a_1,(a_2, \dots (a_{n-2}, a_{n+1})\dots); \dots; a_1, (a_2, a_{2n-3}); a_1, a_{2n-2}; a_{2n-1}}}
\end{equation}
where $a_{2n-1}$ is the global anyonic charge, and the parenthesis shows the splitting order.

\begin{figure}[!hb]
    \centering
    \includegraphics[width=0.35\textwidth]{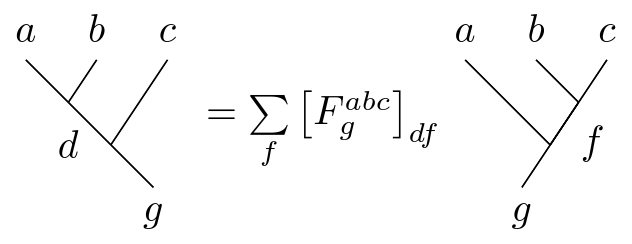}
    \caption{The $F$-matrix defines a change of basis between the different splitting sequences.}
    \label{fig:Fmoves}
\end{figure}

Notice that the splitting diagrams we have chosen to represent the orthonormal basis of the anyonic $N$-particle systems follow a specific splitting order. Changing such order represents choosing a different orthonormal basis of $\pazocal{H}_N$. The change of basis is defined with a set of unitary matrices ($F$-matrices). There, we are moving from basis $\{\ket{(a,b),c;d,c;g}\}$ to basis $\{\ket{a,(b,c); a,f;g}\}$, given by
\begin{align}
 \ket{(a,b),c;d,c;g}=\sum_f [F^{abc}_g]_{df} \ket{a,(b,c); a,f;g}.  
\end{align}
See Fig.~\ref{fig:Fmoves} for diagrammatic representation. Notice that the global or external charges do not change, only the splitting order. In Fibonacci anyons, the only nontrivial $F$-matrix is $F_{\tau}^{\tau \tau \tau}=\begin{pmatrix}
\phi^{-1} & \phi^{-1/2}\\
\phi^{-1/2} & -\phi^{-1}
\end{pmatrix}$.  $\phi^{-1}=\frac{\sqrt{5}-1}{2}$ is the inverse golden ratio. 

\begin{figure}[!ht]
    \centering
    \includegraphics[width=0.25\linewidth]{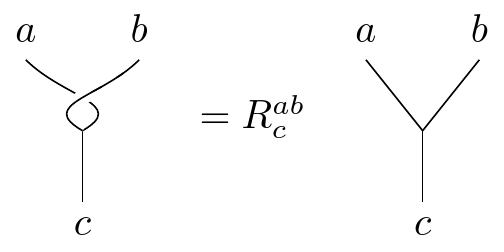}
    \caption{The phase resulting after exchanging counterclockwise two particles $a$ and $b$ that fuse to $c$ is $R_c^{ab}$.}
    \label{fig:braiding}
\end{figure}

The last important element of an anyon theory is the specification of the phase the system gathers when two anyon particles are exchanged. For different anyon families, one can achieve a wide range of exchange phases beyond the acquired phases $+1$ and $-1$ for bosons and fermions, respectively. Thus, the name ``any''on was originated. However, two anyons living in two dimensions can be exchanged either clockwise or counterclockwise. This is represented by the anyon particle lines being able to cross over or underneath each other. 

The $R$-matrices from Fig.~\ref{fig:braiding} encode the exchange phases of the anyons in the theory. The vacuum particle type always gathers the trivial phase $+1$ when exchanged with any other particle type. For Fibonacci anyons, there are two non-trivial $R$-matrix elements: (i) when two $\tau$ anyons are exchanged and fusing to the identity, and (ii) when two $\tau$ anyons are exchanged, fusing to $\tau$. Their respective phases are $R_e^{\tau \tau}=e^{-4\pi i/5}$ and $R_{\tau}^{\tau \tau}=e^{3\pi i/5}$. Such phases play no role in our main text analysis.

\subsection{Superselection rule, observables and density operators}
\label{subsec:operators}

We have presented the basic kinematical structure of non-Abelian anyonic systems. However, such systems incorporate a charge super-selection rule (cSSR) \cite{Simon2023TopologicalQuantum}. The cSSR is analogous to the fermionic parity super-selection rule \cite{Wick1952TheParticles, Friis16, Friis2013Fermionic-modeInformation, Kraus2009PairingPerspective}. By this, the superpositions of states with different total particle types are forbidden. In Fig.~\ref{fig:ket}, only superpositions of states with the same $a_{2n-1}$ can represent physical states. We can refer to anyonic particle types as \emph{charges}. The $a_{2n-1}$ is considered the global charge of the system since, as seen from far away, the system would appear as if all its anyons had fused. 
\begin{figure}[!hb]
    \centering
    \includegraphics[width=0.4\linewidth]{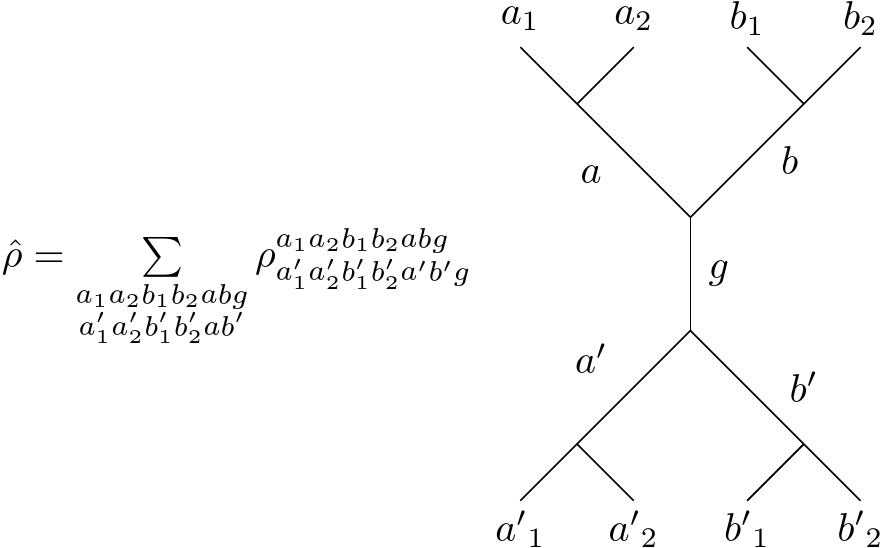}
    \caption{Block-diagonal operators that respect the anyonic SSR.}
    \label{fig:observables}
\end{figure}
The global charge superselection rule (cSSR) induces a block diagonal structure of the kinematical Hilbert spaces $\pazocal{H}_N=\bigoplus_{g} \pazocal{H}_N^{g}$, where $\pazocal{H}_N^g$ is the Hilbert space spanned by the $N$-particle splitting diagrams that have a global anyonic charge $g$. 

Let us now represent operators that act on the Hilbert space $\pazocal{H}_N$ that are block diagonal, respecting the superselection rule structure. Such operators can be written using Dirac notations as sums of $\ketbra{\psi_i}{\psi_j}$ where $\ket{\psi_i}$ and $\bra{\psi_j}$ have the same global charge. Fig.~\ref{fig:observables} represents a general block diagonal operator by stacking the diagram representing $\ket{\psi_i}$ on top of the diagram of $\bra{\psi_j}$. Notice the ket and the bra diagrams can be connected because they have the same global anyonic charge $g$.  

\begin{figure}[!ht]
    \centering
    \includegraphics[width=0.25\linewidth]{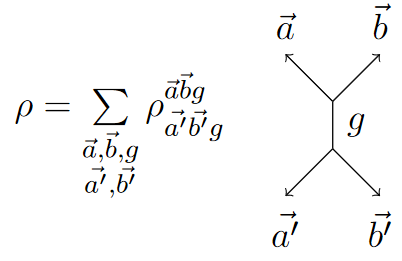}
    \caption{All the particle types of the branches are expressed in the ordered vectors $\vec{a}$, $\vec{b}$, $\vec{a'}$ and $\vec{b'}$.}
    \label{fig:notation}
\end{figure}

Throughout this work, we focus on bipartite non-Abelian anyonic systems. Without loss of generality, we consider systems with $2N$ anyon particles. We choose as a default a splitting basis that showcases the bipartition of the system. In Fig.~\ref{fig:notation}, we introduce useful notation to ease the exposition and calculations. We partition our anyonic system into two regions, $A$ and $B$, representing the global fusion charge's left and right splitting. The vectors $\vec{a}$, $\vec{b}$, $\vec{a'}$, and $\vec{b'}$ encode the information of which particle types appear in the subsequent splitting.  

In the Dirac notation, the cSSR respecting operator in Fig.~\ref{fig:observables} is expressed as 
\begin{equation}
    \Scale[0.96]{\hat{\rho}=\sum \rho_{\vec{a'},\vec{b'},g}^{\vec{a},\vec{b},g} \ketbra{(a_1,a_2),(b_1,b_2);a,b;g}{({a'}_1,{a'}_2),({b'}_1,{b'}_2);a',b';g}}.
\end{equation}

In the Dirac notation, we can use the compression from Fig.~\ref{fig:notation} as well. If we express the terms of Fig.~\ref{fig:observables} in this compressed form we get the identification $\vec{a}=[a,a_2,a_1]$, $\vec{a'}=[a',{a'}_2,{a'}_1]$, $\vec{b}=[b,b_2,b_1]$ and $\vec{b'}=[b',{b'}_2,{b'}_1]$, leading to the identification: 
\begin{eqnarray}\label{eq:compvec}
    &\ketbra{\vec{a},\vec{b};g}{\vec{a'},\vec{b'};g}=\nonumber \\ &=\ketbra{[a,a_2,a_1],[b,b_2,b_1];g}{[a',{a'}_2,{a'}_1],[b',{b'}_2,{b'}_1];g} \equiv  \nonumber \\ &\equiv \ketbra{(a_1,a_2),(b_1,b_2);a,b;g}{({a'}_1,{a'}_2),({b'}_1,{b'}_2);a',b';g}
\end{eqnarray}

\subsection{Trace and partial trace}
\label{subsec:trace}
\begin{figure}[!ht]
    \centering
    \includegraphics[width=0.2\linewidth]{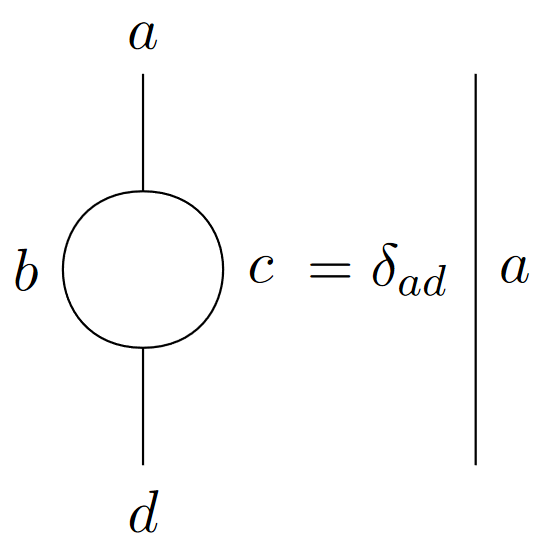}
    \caption{Evaluation of a simple bubble diagram.}
    \label{fig:loop}
\end{figure}

The trace of an operator is inherited from the orthonormality condition of the splitting basis. Thus, $\Tr(\ketbra{\psi_i}{\psi_j})=\braket{\psi_j}{\psi_i}$, with the right-hand side dictated from the orthonormality condition of the splitting diagrams. In the vector notation, we have that $\braket{a,b;g}{a',b';g'}=\delta_{a a'} \delta_{b b'} \delta_{g g'}$. In the general bipartite compressed vector form, we have $\braket{\vec{a}, \vec{b};g}{\vec{a'}, \vec{b'};g'}=\delta_{\vec{a} \vec{a'}} \delta_{\vec{b} \vec{b'}} \delta_{g g'}$. Such relation hints to the diagrammatical equation in Fig.~\ref{fig:loop}. Any single closed loop is evaluated to $1$. 

\begin{figure}[!ht]
    \centering
    \includegraphics[width=0.35\linewidth]{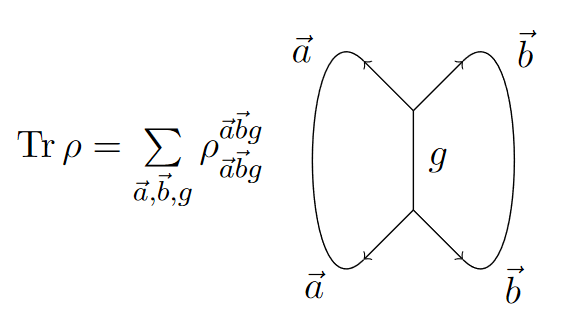}
    \caption{Diagrammatic trace of the SSR operator $\rho$.}
    \label{fig:totaltrace}
\end{figure}

The trace is generally represented diagrammatically as in Fig.~\ref{fig:totaltrace}. These closed-loop diagrams are evaluated by applying $F$ and $R$-matrices to 'untie' the strands to use the relation shown in Fig.~\ref{fig:loop}.  

This trace allows us to define the states of an $N$-mode anyonic system the usual way, as the cSSR respecting operators $\rho$ of $\pazocal{H}_N$ such that $\rho=\rho^\dagger$, $\rho \geq 0$ and $\Tr(\rho)=1$. Nevertheless, it is not enough, we need the notion of partial trace.  

\begin{figure}[!ht]
    \centering
    \includegraphics[width=0.6\linewidth]{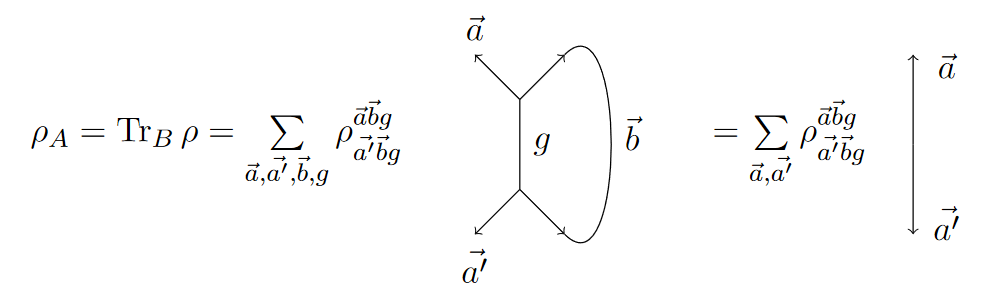}
    \caption{Partial trace of a density matrix $\rho$.}
    \label{fig:partialtrace}
\end{figure}

\begin{figure}[!ht]
    \centering
    \includegraphics[width=0.6\linewidth]{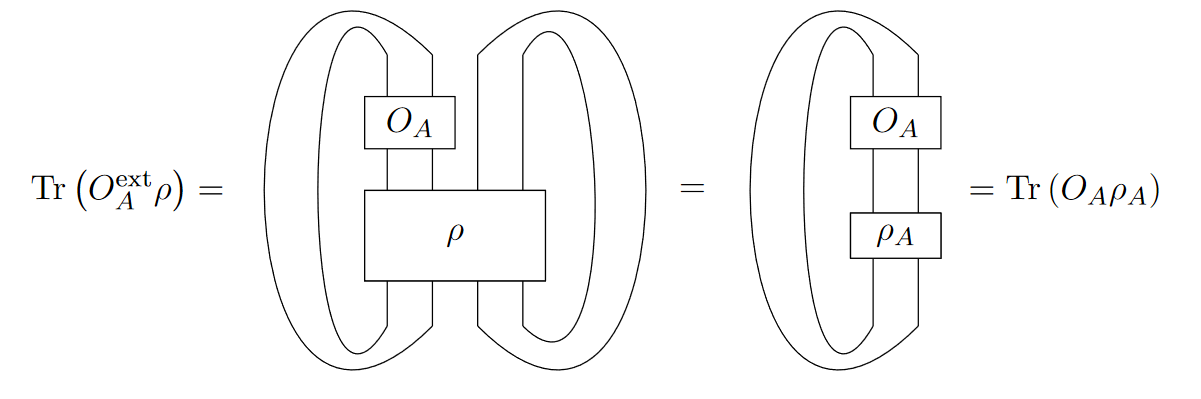}
    \caption{Diagrammatical consistency conditions of the partial trace with anyons.}
    \label{fig:consistency}
\end{figure}

The partial trace of a cSSR respecting operator over $B$ in the bipartite system $AB$ is shown in Fig.~\ref{fig:partialtrace}. It consists of joining the anyon lines in a loop in $B$, computing the relevant factors using the $F$ and $R$ matrices, and the loop evaluations from Fig.~\ref{fig:loop}. It is straightforward to check that such a definition is fixed by the usual defining consistency conditions of the partial trace, which are: 
\begin{equation}
    \Tr\left(\hat{O}_A \cdot \Tr_B\left(\rho_{AB}\right)\right)=\Tr\left(\hat{O}_A \ \rho_{AB}\right); \quad \forall \ \hat{O}_A \in \pazocal{O}_A,
\end{equation}
where $\pazocal{O}_A$ is the set of all cSSR respecting local observables belonging to subsystem $A$. Fig.~\ref{fig:consistency} shows diagrammatically how the partial trace satisfies the consistency conditions.

Reading the diagrammatic partial trace over $B$ in the compressed notation in Fig.~(\ref{fig:totaltrace}), we obtain the following expression, where $a_0,{a'}_0$ are the first terms of the vectors $\vec{a}$ and $\vec{a'}$, and thus the global charges of $A$ for the 'ket' and 'bra', i.e., 
\begin{eqnarray}\label{eq:vecpt}
    \Tr_B\left(\ketbra{\vec{a}, \vec{b};g}{\vec{a'}, \vec{b'};g}\right)= \delta_{\vec{b} \vec{b'}} \delta_{a_0 {a'}_0} \ketbra{\vec{a}}{\vec{a'}},
\end{eqnarray}
where $\ketbra{\vec{a}}{\vec{a'}}$ denotes the anyonic operator term in $A$ in the specified splitting order with the particle types described by the vectors $\vec{a},\vec{a'}$ components. This general expression implies the ones presented in the main text.

\section{Purity of non-Abelian anyonic states}
 \label{sec:appendix2}

In quantum mechanics, a pure state is a state that is represented by a single ket or a coherent superposition of kets in the corresponding vector space. From the perspective of the geometry of quantum state space, a quantum state $\rho$ is pure if it cannot be written as a non-trivial convex combination of two distinct states. The states that are not pure are called mixed. 
A usual measure of the purity of a given quantum state $\rho$ is given by 
\begin{equation}
\gamma = \Tr(\rho^2). 
\end{equation}
If the state $\rho$ is pure, then the purity of the state $\gamma =1$. For a maximally mixed state $\gamma = \frac{1}{d}$ where $d$ is the dimension of the Hilbert space. 

In previous works that incorporated quantum information notions to the study of non-Abelian anyons~\cite{Bonderson2017AnyonicEntropy}, the authors claim that all states with a non-Abelian anyonic global charge are not pure. This is due to the way they define non-Abelian anyonic states. They argue that anyons with non-trivial charge cannot exist physically on their own without a corresponding anti-anyon, such that the global charge would be $e$. Thus, any non-trivial charged anyonic state must be derived by tracing out the anti-anyon. This results in anyonic states with charge $a$ to have factor of $\frac{1}{\sqrt{d_a}}$ with the corresponding anyon diagram, where $d_a$ is the quantum dimension of the anyon. This factor happens to be non-trivial for non-Abelian anyons.  This approach is useful when doing topological manipulations of the anyonic particles, as it preserves the isotopy invariance of the anyonic state~\cite{Bonderson2017AnyonicEntropy}. We call this normalization the `isotopy invariant normalization.' 

If one calculates $\gamma$ using the isotopy invariant normalisation, one obtains that for the single Fibonacci state $\ketbra{\tau}$, the purity $\gamma=\frac{1-\sqrt{5}}{2}<1$. Therefore, it is claimed that $\ketbra{\tau}$ is not pure. In general, according to the isotopy invariant normalization, no state with non-Abelian anyonic total charge can be considered a pure state. This is obviously in contention with the normalization that we use throughout this paper which allows for states with global charge $\tau$ to be pure.

We use a proof similar to the case of qudit systems \cite{Nielsen00} to argue that states with non-Abelian anyonic global charge can be pure states. Since anyonic systems can be represented by states in a Hilbert space, we use the definition of pure states for a convex vector space.  Assume that $\ketbra{\psi_1}$ is not a pure state. Then, there exist distinct states $\sigma, \omega$ and a probability $p\in (0,1)$ such that $\ketbra{\psi_1}=p \sigma +(1-p) \omega$.    Let us consider $\ketbra{\psi_1}$, $\omega$ and $\sigma$ as unit vectors of the operator Hilbert space, that has as scalar product $\langle \hat{A}, \hat{B}\rangle \equiv \Tr(\hat{A}^\dagger \hat{B})$. Consider the orthonormal basis in such operator vector space $\{\ketbra{\psi_k}{\psi_l}\}_{k,l}$, whose first element is $\ketbra{\psi_1}$. Let us take the first coordinate of $\ketbra{\psi_1}$ in such a basis. To do so, let us calculate $\langle \ketbra{\psi_1}, \ketbra{\psi_1}\rangle$, where the first is the vector and the second as the first element of the orthonormal basis. Then, we obtain

\begin{align}
    \langle \ketbra{\psi_1}, \ketbra{\psi_1}\rangle = \Tr( \ketbra{\psi_1} \cdot \ketbra{\psi_1})=1= p\Tr(\sigma \cdot \ketbra{\psi_1}) + (1-p)\Tr(\omega \cdot \ketbra{\psi_1})= \nonumber \\ =p \bra{\psi_1} \sigma \ket{\psi_1} +(1-p) \bra{\psi_1} \omega \ket{\psi_1}
\end{align}

Now, since $\sigma$ and $\omega$ are physical states, by definition, they must be positive operators with trace 1. Therefore $0 \leq\bra{\psi_1} \sigma \ket{\psi_1} \leq 1$ and $0 \leq\bra{\psi_1} \omega \ket{\psi_1} \leq 1$. Since $p\in (0,1)$, the only way to satisfy the equality is by having $\bra{\psi_1} \sigma \ket{\psi_1} = 1$ and $\bra{\psi_1} \omega \ket{\psi_1}= 1$. However, then since the trace is $1$, we obtain  $\bra{\psi_k} \sigma \ket{\psi_k} = 0$ and $\bra{\psi_k} \omega \ket{\psi_k}= 0$ for all $k >1$. Since we need $\sigma$ and $\omega$ to be positive and hermitian operators, this implies that then $\bra{\psi_k} \sigma \ket{\psi_1} = 0$ and $\bra{\psi_k} \omega \ket{\psi_1}= 0$ for $k>1$. Therefore, we obtain that $\omega=\sigma=\ketbra{\psi_1}$ leading to a contradiction with the assumption that $\sigma$ and $\omega$ were distinct states. This implies that $\ketbra{\psi_1}$ must necessarily be a pure state in the anyonic Hilbert space. Note that we made no assumptions about the $\ketbra{\psi_1}$ total charge in this analysis; thus, this result must hold even for states with non-Abelian anyonic total charge. Thus, any anyonic state that can be expressed as a ket can be considered to be a pure state.

A summarising comment would be that the notion of purity of anyonic states must not be considered in the isotopy invariant normalization for information-theoretic analysis since the formulas for purity and other information-theoretic notions rely on having the standard normalization $\braket{\psi}=1$ for physical states.

\section{Uncorrelated states for a 2-anyon system}
\label{sec:appendixuncorr}
We have claimed in the main text that the pure uncorrelated states in a 2-anyon Fibonacci system are of the form $\alpha_e=0$ or $\beta_e=0$ in the $e$ sector and $\alpha_\tau=0$ or $\beta_\tau=0$ in the $\tau$ sector. The coefficients $\alpha_e,\beta_e$ refer to Eq.~(\ref{eq:estates}) and $\alpha_\tau,\beta_\tau,\gamma_\tau$ refer to Eq.~(\ref{eq:taustates}). In this Appendix, we explicitly show the calculations that prove these results.  

We need to solve Eq.~(\ref{eq:uncorr}) for general $\rho_{AB}=\ketbra{\psi_e}_{AB} \equiv \varepsilon_{AB}$ and $\rho_{AB}=\ketbra{\psi_\tau}_{AB}\equiv \sigma_{AB}$. To do so, it is straightforward to see that any local observable in the 1-anyon Fibonacci systems $A$ and $B$ can be represented in the basis $\{\ket{e},\ket{\tau}\}$ as matrices
\begin{eqnarray}
    \hat{O}_A=\begin{pmatrix}
        a_e & 0 \\ 0 & a_\tau
    \end{pmatrix}, \ \  \hat{O}_B=\begin{pmatrix}
        b_e & 0 \\ 0 & b_\tau
    \end{pmatrix} \quad \text{with} \quad  a_e, a_\tau, b_e, b_\tau \in \mathbb{R}~.
\end{eqnarray}
The local observables $\hat{O}_A$ and $\hat{O}_B$ embedded in the 2-anyon $AB$ system can be represented in the basis $\{\ket{e,e:e}, \ket{\tau,\tau;e}, \ket{\tau,e;\tau}, \ket{e,\tau;\tau}, \ket{\tau,\tau;\tau}\}$ as the matrices
\begin{eqnarray}
    \hat{O}_A=\begin{pmatrix}
        a_e & 0 & 0 & 0 & 0 \\ 0 & a_\tau & 0 & 0 & 0 \\0 & 0 &  a_\tau & 0 & 0 \\ 0 & 0 & 0 & a_e & 0 \\ 0 & 0 & 0 & 0 & a_\tau  
    \end{pmatrix} \qquad \mbox{and} \qquad \hat{O}_B=\begin{pmatrix}
        b_e & 0 & 0 & 0 & 0 \\ 0 & b_\tau & 0 & 0 & 0 \\0 & 0 &  b_e & 0 & 0 \\ 0 & 0 & 0 & b_\tau & 0 \\ 0 & 0 & 0 & 0 & b_\tau  
    \end{pmatrix} ~.
\end{eqnarray}

Applying the anyonic partial trace from Eq.~(\ref{eq:pt2}), we obtain that the matrix representations of $\varepsilon_A, \varepsilon_B, \sigma_A, \sigma_B$ in the 1-anyon basis we have used above are
\begin{eqnarray}
    \varepsilon_{A}=\begin{pmatrix}
        |\alpha_e|^2 & 0 \\ 0 & |\beta_e|^2
    \end{pmatrix}= \varepsilon_B, \qquad \sigma_{B}=\begin{pmatrix}
        |\alpha_\tau|^2 & 0 \\ 0 & |\beta_\tau|^2+|\gamma_\tau|^2
    \end{pmatrix}, \medspace \medspace \sigma_{A}=\begin{pmatrix}
        |\beta_\tau|^2 & 0 \\ 0 & |\alpha_\tau|^2+|\gamma_\tau|^2
    \end{pmatrix} ~.
\end{eqnarray}

Thus, solving Eq.~(\ref{eq:uncorr}) for $\varepsilon_{AB}$ is equivalent to equating and solving $\forall a_e, a_\tau, b_e, b_\tau \in \mathbb{R}$ the expressions
\begin{align}
    & \Tr\left( \begin{pmatrix}
        a_e & 0 & 0 & 0 & 0 \\ 0 & a_\tau & 0 & 0 & 0 \\0 & 0 &  a_\tau & 0 & 0 \\ 0 & 0 & 0 & a_e & 0 \\ 0 & 0 & 0 & 0 & a_\tau  
    \end{pmatrix} \cdot \begin{pmatrix}
        b_e & 0 & 0 & 0 & 0 \\ 0 & b_\tau & 0 & 0 & 0 \\0 & 0 &  b_e & 0 & 0 \\ 0 & 0 & 0 & b_\tau & 0 \\ 0 & 0 & 0 & 0 & b_\tau  
    \end{pmatrix} \cdot \begin{pmatrix}
        |\alpha_e|^2 & \alpha_e \beta_e^* & 0 & 0 & 0 \\ \beta_e \alpha_e^*  & | \beta_e|^2  & 0 & 0 & 0 \\0 & 0 &  0 & 0 & 0 \\ 0 & 0 & 0 & 0 & 0 \\ 0 & 0 & 0 & 0 & 0  
    \end{pmatrix} \right) = a_e b_e |\alpha_e|^2 + a_\tau b_\tau | \beta_e|^2  \\  & \Tr\left( \begin{pmatrix}
        a_e & 0 \\ 0 & a_\tau
    \end{pmatrix} \cdot \begin{pmatrix}
        |\alpha_e|^2 & 0 \\ 0 & |\beta_e|^2
    \end{pmatrix} \right) \cdot \Tr\left(\begin{pmatrix}
        b_e & 0 \\ 0 & b_\tau
    \end{pmatrix} \cdot \begin{pmatrix}
        |\alpha_e|^2 & 0 \\ 0 & |\beta_e|^2
    \end{pmatrix} \right)=(a_e |\alpha_e|^2+ a_\tau |\beta_e|^2 )\cdot (b_e |\alpha_e|^2+ b_\tau |\beta_e|^2 ).
\end{align}
Choosing the special case where $a_e=1,a_\tau=0,b_e=0,b_\tau=1$ we obtain that the equality $0=|\alpha_e|^2 |\beta_e|^2$ must hold, implying that necessarily either $\alpha_e=0$ or $\beta_e=0$. It is straightforward to see from the equations above and using $|\alpha_e|^2 +|\beta_e|^2=1$ that these two cases are sufficient to guarantee that the state is uncorrelated. Therefore, we recover the announced result. 

Let us now reproduce the same logic for the $\tau$ sector. To solve Eq.~(\ref{eq:uncorr}) for $\sigma_{AB}$ is equivalent to equating and solving $\forall a_e, a_\tau, b_e, b_\tau \in \mathbb{R}$ the expressions
\begin{eqnarray}
    \Scale[0.95]{\Tr\left( \begin{pmatrix}
        a_e & 0 & 0 & 0 & 0 \\ 0 & a_\tau & 0 & 0 & 0 \\0 & 0 &  a_\tau & 0 & 0 \\ 0 & 0 & 0 & a_e & 0 \\ 0 & 0 & 0 & 0 & a_\tau  
    \end{pmatrix} \cdot \begin{pmatrix}
        b_e & 0 & 0 & 0 & 0 \\ 0 & b_\tau & 0 & 0 & 0 \\0 & 0 &  b_e & 0 & 0 \\ 0 & 0 & 0 & b_\tau & 0 \\ 0 & 0 & 0 & 0 & b_\tau  
    \end{pmatrix} \cdot \begin{pmatrix}
        0 & 0  & 0 & 0 & 0 \\ 0  & 0  & 0 & 0 & 0 \\0 & 0 &  |\alpha_\tau|^2 & \alpha_\tau \beta_\tau^*  & \alpha_\tau \gamma_\tau^* \\ 0 & 0 & \beta_\tau \alpha_\tau^* & |\beta_\tau|^2 & \beta_\tau \gamma_\tau^* \\ 0 & 0 & \gamma_\tau \alpha_\tau^* & \gamma_\tau \beta_\tau^* & |\gamma_\tau|^2  
    \end{pmatrix} \right) = a_\tau b_e |\alpha_\tau|^2 + a_e b_\tau | \beta_\tau|^2 + a_\tau b_\tau | \gamma_\tau|^2}  \\ \Tr\left( \begin{pmatrix}
        a_e & 0 \\ 0 & a_\tau
    \end{pmatrix} \cdot \begin{pmatrix}
        |\beta_\tau|^2 & 0 \\ 0 & |\alpha_\tau|^2+|\gamma_\tau|^2 
    \end{pmatrix} \right) \cdot \Tr\left(\begin{pmatrix}
        b_e & 0 \\ 0 & b_\tau
    \end{pmatrix} \cdot \begin{pmatrix}
        |\alpha_\tau|^2 & 0 \\ 0 & |\beta_\tau|^2+|\gamma_\tau|^2
    \end{pmatrix} \right)= \nonumber \\ =(a_e |\beta_\tau|^2+ a_\tau |\alpha_\tau|^2+ a_\tau |\gamma_\tau|^2 )\cdot (b_e |\alpha_\tau|^2+ b_\tau |\beta_\tau|^2 +b_\tau |\gamma_\tau|^2 ).
\end{eqnarray}
Similarly, choosing the special case where $a_e=1,a_\tau=0,b_e=1,b_\tau=0$ we obtain that the equality $0=|\alpha_\tau|^2 |\beta_\tau|^2$ must hold, implying that necessarily either $\alpha_\tau=0$ or $\beta_\tau=0$. To show the sufficiency of these conditions, let us see if the equality between the expressions above holds whenever $\alpha_\tau=0$ or $\beta_\tau=0$. One can observe that the two cases are symmetric. So let us fix without loss of generality that $\alpha_\tau=0$. Notice that in such case $|\beta_\tau|^2+|\gamma_\tau|^2=1$. Then, the expression equality becomes
\begin{eqnarray}
    b_\tau \left(a_e |\beta_\tau|^2+ a_\tau|\gamma_\tau|^2\right)= \left(a_e |\beta_\tau|^2 +a_\tau |\gamma_\tau|^2 \right) \left(b_\tau |\beta_\tau|^2+b_\tau |\gamma_\tau|^2\right)
\end{eqnarray}
The right-hand side of the equation can be simplified by taking $b_\tau$ as a common factor and then using $|\beta_\tau|^2+|\gamma_\tau|^2=1$ to obtain exactly the same expression as in the left-hand side. Thus proving the sufficiency of the $\alpha_\tau=0$ condition for having a pure uncorrelated state. As we said, the case for $\beta_\tau=0$ follows from symmetry in an exactly equal manner.  

\section{More teleportation examples}
\label{sec:tpexamples}

In this section, we will look at the teleportation protocol between Alice and Bob, using the same quantum message state as in the main text 

\begin{align}
    \ket{\varphi_M} &= \alpha\ket{\tau,e;\tau} + \beta\ket{e,\tau;\tau},
\end{align}
but with different resource states. Below we show that whenever the resource entangled state, shared between Alice and Bob, has global charge $e$, the quantum teleportation is symmetric. However, a resource entangled state with a global charge $\tau$ may lead to an asymmetric quantum teleportation.  

\subsection{Entangled state in the e-charge sector: symmetric teleportation}
We begin our protocol by considering the following state as the resource state
\begin{align}
    \ket{R_{AB}} = \frac{1}{\sqrt{2}}\Big(\ket{(e,e),(e,e);e,e;e} + \ket{(\tau,\tau),(\tau,\tau);e,e;e}\Big).
    \label{eq:AB_e}
 \end{align}
The global state of Alice and Bob is
 \begin{align}
     |\psi_{M(AB)}\rangle = \frac{1}{\sqrt{2}}\Big(&\alpha\ket{(\tau,e),(e,e),(e,e);\tau,(e,e);\tau,e;\tau} + \beta\ket{(e,\tau),(\tau,\tau),(\tau,\tau);\tau,(e,e);\tau,e;\tau}\nonumber\\  + &\alpha\ket{(\tau,e),(\tau,\tau),(\tau,\tau);\tau,(e,e);\tau,e;\tau} + \beta\ket{(e,\tau),(e,e),(e,e);\tau,(e,e);\tau,e;\tau}\Big).
     \label{eq:CAB_e}
 \end{align}
Here, we can re-express the state combining $MA$, as defined in Eq.~\eqref{eq:AlicePVM_e},
\begin{align*}
     |\psi_{(MA)B}\rangle = \frac{1}{\sqrt{2}}\Big(\alpha(\ket{\lambda_+}+\ket{\lambda_-})\ket{e,e;e} + \beta(\ket{\lambda_+}-\ket{\lambda_-})\ket{\tau,\tau;e}\\  + \alpha(\ket{\theta_+}+\ket{\theta_-})\ket{\tau,\tau;e} + \beta(\ket{\theta_+}-\ket{\theta_-})\ket{e,e;e}\Big),
 \end{align*}
where 
\begin{align}
     \ket{\lambda_{\pm}}= \frac{1}{\sqrt{2}} \Big( \ket{(\tau,e),(e,e);\tau,e;\tau} \pm  \ket{(e,\tau),(\tau,\tau);\tau,e;\tau} \Big) \nonumber \\
     \ket{\theta_{\pm}}= \frac{1}{\sqrt{2}} \Big( \ket{(\tau,e),(\tau,\tau);\tau,e;\tau} \pm \ket{(e,\tau),(e,e);\tau,e;\tau} \Big),
     \label{eq:AlicePVM_e}
\end{align}
are orthonormal states. 

The Alice to Bob teleportation protocol will then proceed as follows. Alice performs measurements with the projectors $\{\ketbra{\lambda_+}, \ketbra{\lambda_-}, \ketbra{\theta_+}, \ketbra{\theta_-} \} $ and communicates the result over a classical channel whenever any of these projective measurement clicks, and Bob can then apply local cSSR-respecting unitary operations $\{X,Y,I,Z\}$ respectively, where $Z=\ketbra{\tau,\tau;e}-\ketbra{e,e;e}$ and recover the quantum message. This is an instance of perfect teleportation between Alice and Bob. 
 
 The resource state in Eq.~\eqref{eq:AB_e} is symmetric between Alice and Bob. It is thus easy to verify that for teleportation from Bob to Alice. The global state is then
 \begin{align}
     \ket{\phi_{(AB)M}} = \frac{1}{\sqrt{2}}\Big(&\alpha\ket{(e,e),(e,e),(\tau,e);(e,e),\tau;\tau,e;\tau} + \beta\ket{(\tau,\tau),(\tau,\tau),(e,\tau);(e,e),\tau;e,\tau;\tau}\nonumber\\  + &\alpha\ket{(\tau,\tau),(\tau,\tau),(\tau,e);(e,e),\tau;e,\tau;\tau} + \beta\ket{(e,e),(e,e),(e,\tau);(e,e),\tau;e,\tau;\tau}\Big),
 \end{align}
 which differs from the state in Eq.~\eqref{eq:CAB_e} by a $F$-move, i.e. a change of basis. However, the F-move will not introduce any non-trivial phases. It follows that Bob will be able to apply the same measurement as Alice and that the probability of successful teleportation would be the same in either case. It can be seen easily that the resource state in Eq.~\eqref{eq:AB_e} is symmetric with respect to the parties $A$ and $B$. This would be the case for any $4$-anyon bipartite state where each of the subsystems has a fixed charge of $e$, as the only possible candidates for such (normalized) states will have the form
 \begin{align}
    \ket{R'_{AB}} = \Big(a\ket{(e,e),(e,e);e,e;e} + b\ket{(\tau,\tau),(\tau,\tau);e,e;e}\Big). 
    \label{eq:AB'_e}
 \end{align}
 Since, all such states are symmetric between $A$ and $B$, the teleportation protocol would also be symmetric, i.e, the teleportation fidelity would be independent of the direction of communication.\\

 \subsection{Entangled state with asymmetric marginals: asymmetric teleportation }
 Let us consider the teleportation of the quantum message $\ket{\varphi_M}$, using a resource entangled state shared between Alice and Bob, given by
 \begin{align}
 \label{eq:AB_asym}
     \ket{R_{AB}} = \frac{1}{2}\Big(\sqrt{2}\ket{(e,e),(e,\tau);e,\tau;\tau} + \ket{(e,\tau),(e,e);\tau,e;\tau} + \ket{(\tau,e),(e,\tau);\tau,\tau;\tau}\Big). 
\end{align}
When taking the partial trace of the above state for each of the parties, the reduced states become 
\begin{align*}
\rho_A =
    \begin{pmatrix}
        \frac{1}{2} & 0 & 0 & 0 & 0\\
        0 & 0 & 0 & 0 & 0 \\
        0 & 0 & \frac{1}{4} & 0 & 0 \\
        0 & 0 & 0 & \frac{1}{4} & 0 \\
        0 & 0 & 0 & 0 & 0 \\
    \end{pmatrix}
    \quad \mbox{and } \quad
 \rho_B =
 \begin{pmatrix}
     \frac{1}{4} & 0 & 0 & 0 & 0 \\
     0 & 0 & 0 & 0 & 0 \\
     0 & 0 & 0 & 0 & 0 \\
     0 & 0 & 0 & \frac{3}{4} & 0 \\
     0 & 0 & 0 & 0 & 0 \\
 \end{pmatrix}.
\end{align*}
By comparing the reduced density matrices (on the standard basis), we see that they have asymmetry in their marginal spectra.
First, let us consider the case where Alice teleports the state to Bob. The global state of Alice and Bob is now
\begin{align}
    |\psi_{M(AB)}\rangle = \frac{1}{2}( \sqrt{2}&\alpha\ket{(\tau,e),(e,e),(e,\tau);\tau,(e,\tau);\tau,\tau;e}+\sqrt{2}\beta\ket{(e,\tau),(e,e),(e,\tau);\tau,(e,\tau);\tau,\tau;e}\nonumber\\ + &\alpha\ket{(\tau,e),(e,\tau),(e,e);\tau,(\tau,e);\tau,\tau;e} + \beta\ket{(e,\tau),(e,\tau),(e,e);\tau,(\tau,e);\tau,\tau;e}\nonumber\\ + &\alpha\ket{(\tau,e),(\tau,e),(e,\tau);\tau,(\tau,\tau);\tau,\tau;e} + \beta\ket{(e,\tau),(\tau,e),(e,\tau);\tau,(\tau,\tau);\tau,\tau;e} ).
\end{align}
Here, After an F-move, we can rewrite the states as
\begin{align}
    \ket{\psi_{(MA)B}} = &\frac{1}{2}\Big( \big(\sqrt{2}\alpha\ket{(\tau,e),(e,e);\tau,e;\tau} +\sqrt{2}\beta\ket{(e,\tau),(e,e);\tau,e;\tau}\nonumber\\ & + \alpha\ket{(\tau,e),(\tau,e);\tau,\tau;\tau} + \beta\ket{(e,\tau),(\tau,e);\tau,\tau;\tau}\big)\ket{e,\tau;\tau}\nonumber \\ &+ \big(\alpha\ket{(\tau,e),(e,\tau);\tau,\tau;e} + \beta\ket{(e,\tau),(e,\tau);\tau,\tau;e}\big)\ket{e,e;e}\Big).
\end{align}
We have also factored out the $B$ state wherever possible. We will now re-express the state as 
\begin{align}
    \ket{\psi_{(MA)B}} = \frac{1}{2}\Big( \ket{\mu;\tau}\ket{e,\tau;\tau} + \ket{\nu;e}\ket{e,e;e}\Big),
    \label{eq:asymfactorizedCAB}
\end{align}
where we denote
\begin{align}
    \ket{\mu;\tau} &= \sqrt{2}\alpha\ket{(\tau,e),(e,e)\tau,e;\tau} +\sqrt{2}\beta\ket{(e,\tau),(e,e);\tau,e;\tau}\nonumber\\ &+ \alpha\ket{(\tau,e),(\tau,e);\tau,\tau;\tau} + \beta\ket{(e,\tau),(\tau,e);\tau,\tau;\tau} \\
    \ket{\nu;e} &= \alpha\ket{(\tau,e),(e,\tau);\tau,\tau;e}+ \beta\ket{(e,\tau),(e,\tau);\tau,\tau;e}.
\end{align}
Any projective measurement by Alice on the $MA$ system can, at best, result in Bob's local state being a classical mixture of $\ket{e,\tau;\tau}$ and $\ket{e,e;e}$, since these two states cannot be superposed due to cSSR. Such teleportation can be achieved by using a classically correlated state. Thus, there cannot be any quantum teleportation from Alice to Bob.\\ 

For Bob to Alice teleportation, the global state of the $(AB)M$ system will be
\begin{align}
    \ket{\psi_{(AB)M}} = \frac{1}{2}\Big( \sqrt{2}&\alpha\ket{(e,e),(e,\tau),(\tau,e);(e,\tau),\tau;\tau,\tau;e}+\sqrt{2}\beta\ket{(e,e),(e,\tau),(e,\tau);(e,\tau),\tau;\tau,\tau;e}\nonumber\\ + &\alpha\ket{(e,\tau),(e,e),(\tau,e);(\tau,e),\tau;\tau,\tau;e} + \beta\ket{(e,\tau),(e,e),(e,\tau);(\tau,e),\tau;\tau,\tau;e}\nonumber\\ + &\alpha\ket{(\tau,e),(e,\tau),(\tau,e);(\tau,\tau),\tau;\tau,\tau;e} + \beta\ket{(\tau,e),(e,\tau),(e,\tau);(\tau,\tau),\tau;\tau,\tau;e} \Big).
\end{align}
After F-move, the state can be written as
\begin{align}
    \ket{\omega_{A(BM)}} = \frac{1}{2}\Big( \sqrt{2}&\alpha\ket{(e,e),(e,\tau),(\tau,e);e,(\tau,\tau);\tau,\tau;e}+\sqrt{2}\beta\ket{(e,e),(e,\tau),(e,\tau);e,(\tau,\tau);\tau,\tau;e}\nonumber\\ + &\alpha\ket{(e,\tau),(e,e),(\tau,e);\tau,(e,\tau);\tau,\tau;e} + \beta\ket{(e,\tau),(e,e),(e,\tau);\tau,(e,\tau);\tau,\tau;e}\nonumber\\ + &\alpha\ket{(\tau,e),(e,\tau),(\tau,e);\tau,(\tau,\tau);\tau,\tau;e} + \beta\ket{(\tau,e),(e,\tau),(e,\tau);\tau,(\tau,\tau);\tau,\tau;e} \Big).
\end{align}
We rewrite this state again as
\begin{align}
    \ket{\omega_{A(BM)}} = \frac{1}{2}\Big( \sqrt{2}&\alpha\ket{(e,e),(e,\tau),(\tau,e);e,(\tau,\tau);\tau,\tau;e}+\sqrt{2}\beta\ket{(e,e),(e,\tau),(e,\tau);e,(\tau,\tau);\tau,\tau;e}\nonumber\\ + &\alpha\ket{e,\tau;\tau}(\ket{\lambda_+ + \lambda_-} + \beta\ket{e,\tau;\tau}(\ket{\eta_+}-\ket{\eta_-})\nonumber\\ + &\alpha\ket{\tau,e;\tau}(\ket{\eta_+}+\ket{\eta_-}) + \beta\ket{\tau,e;\tau}(\ket{\lambda_+}-\ket{\lambda_-}) \Big),
\end{align}
where we denote
\begin{eqnarray}
     \ket{\lambda_{\pm}}= \frac{1}{\sqrt{2}} \Big( \ket{(e,e),(\tau,e);e,\tau;\tau} \pm  \ket{(e,\tau),(e,\tau);\tau,\tau;\tau} \Big) \nonumber \\
     \ket{\eta_{\pm}}= \frac{1}{\sqrt{2}} \Big( \ket{(e,\tau),(\tau,e);\tau,\tau;\tau} \pm  \ket{(e,e),(e,\tau);e,\tau;\tau} \Big).
     \label{eq:asymmetricPVM}
\end{eqnarray}
Bob now makes a projective measurement $\{\ketbra{\lambda_+}, \ketbra{\lambda_-}, \ketbra{\eta_+}, \ketbra{\eta_-} \} $. However, it can be seen that Bob will not register a click for these measurements all the time. There will be a click only when the total charge of the $BM$ system is $\tau$. The probability of Bob registering a click is
$$\frac{2|\alpha|^2}{4} + \frac{2|\beta|^2}{4} = \frac{1}{2}.$$
If Bob registers a click after the measurement, then the protocol proceeds as usual, where the outcome of measurement is communicated classically to Alice, who then applies the corresponding unitary from $\{X, Y, I, Z\}$, where $Z=\ketbra{\tau,e;\tau}-\ketbra{e,\tau;\tau}$. However, if Bob doesn't register a click, it implies that the charge of the $BM$ will be $e$, and that Alice will end up with the state $\ketbra{e,e;e}$. Alice has no way of recovering the quantum message from here, and thus the protocol fails. Thus the Bob to Alice teleportation protocol will succeed only half the time. However, this case is still better than the Alice to Bob teleportation protocol for the same resource state in Eq.~\eqref{eq:AB_asym} discussed above. 

This result showcases that the asymmetry in teleportation cannot be removed by considering a state with asymmetric marginals, further reinforcing that the entanglement asymmetry is independent of asymmetry in the marginal spectra.
\end{document}